\documentclass[preprint,12pt]{elsarticle}
\biboptions{numbers,sort&compress}
\newcommand{\reffig}[1]{Fig. \ref{#1}}
\newcommand{\reftab}[1]{Table \ref{#1}}
\newcommand{\refequ}[1]{(\ref{#1})}
\newcommand{\tabincell}[2]{\begin{tabular}{@{}#1@{}}#2\end{tabular}}
\DeclareMathAlphabet{\mathscr}{OT1}{pzc}{m}{it}
\usepackage{threeparttable}
\usepackage{graphicx}
\usepackage{amssymb}
\usepackage{upgreek}
\usepackage{siunitx}
\usepackage{subfigure}
\usepackage{color, xcolor}
\usepackage{caption}
\usepackage{lineno}
\usepackage{amsmath}
\usepackage{threeparttable}
\usepackage{multirow}
\usepackage{marvosym}
\usepackage{makecell}
\usepackage{nomencl} 
\usepackage{ifthen}
\usepackage{soul}
\usepackage{adjustbox}
\renewcommand{\nomgroup}[1]{%
	\item[\textbf{%
		\ifthenelse{\equal{#1}{P}}{Parameters and variables}{}%
		\ifthenelse{\equal{#1}{S}}{Superscripts and subscripts}{}
	}]%
}
\makenomenclature
\journal{International Journal of Hydrogen Energy}

\makeatletter
\def\@author#1{\g@addto@macro\elsauthors{\normalsize%
		\def\baselinestretch{1}%
		\upshape\authorsep#1\unskip\textsuperscript{%
			\ifx\@fnmark\@empty\else\unskip\sep\@fnmark\let\sep=,\fi
			\ifx\@corref\@empty\else\unskip\sep\@corref\let\sep=,\fi
		}%
		\def\authorsep{\unskip,\space}%
		\global\let\@fnmark\@empty
		\global\let\@corref\@empty  %% Added
		\global\let\sep\@empty}%
	\@eadauthor={#1}
}
\makeatother

\begin{document}
\captionsetup[figure]{name={Fig.},labelsep=period}
\begin{frontmatter}

%% Title, authors and addresses

\title{Design of the PID temperature controller for an alkaline electrolysis system with time delays}

%% use the tnoteref command within \title for footnotes;
%% use the tnotetext command for the associated footnote;
%% use the fnref command within \author or \address for footnotes;
%% use the fntext command for the associated footnote;
%% use the corref command within \author for corresponding author footnotes;
%% use the cortext command for the associated footnote;
%% use the ead command for the email address,
%% and the form \ead[url] for the home page:
%%
%% \title{Title\tnoteref{label1}}
%% \tnotetext[label1]{}
%% \author{Name\corref{cor1}\fnref{label2}}
%% \ead{email address}
%% \ead[url]{home page}
%% \fntext[label2]{}
%% \cortext[cor1]{}
%% \address{Address\fnref{label3}}
%% \fntext[label3]{}

%% use optional labels to link authors explicitly to addresses:
%% \author[label1,label2]{<author name>}
%% \address[label1]{<address>}
%% \address[label2]{<address>}
\author[label1]{Ruomei Qi \corref{cor0}}
\author[label1]{Jiarong Li \corref{cor0}}
\author[label1,label2]{Jin Lin \corref{cor1}}
\cortext[cor0]{These authors contributed equally.}
\cortext[cor1]{Corresponding author}
\ead{linjin@tsinghua.edu.cn}
\author[label1,label3]{Yonghua Song}
\author[label4,label5]{Jiepeng Wang}
\author[label5]{Qiangqiang Cui}
\author[label6]{Yiwei Qiu}
\author[label2]{Ming Tang}
\author[label2]{Jian Wang}
\address[label1]{State Key Laboratory of Control and Simulation of Power Systems and Generation
Equipment, Department of Electrical Engineering, Tsinghua University, Beijing, China}
\address[label2]{Tsinghua-Sichuan Energy Internet Research Institute, Chengdu, China}
\address[label3]{Department of Electrical and Computer Engineering, University of Macau, Macau, China}
\address[label4]{School of Materials Science and Engineering, Shanghai University, Shanghai, China}
\address[label5]{Purification Equipment Research Institute of CSIC, Handan, China}
\address[label6]{College of Electrical Engineering, Sichuan University, Chengdu, China}

\begin{abstract}
%% Text of abstract
Electrolysis systems use proportional–integral–derivative (PID) temperature controllers to maintain stack temperatures around set points. However, heat transfer delays in electrolysis systems cause manual tuning of PID temperature controllers to be time-consuming, and temperature oscillations often occur. This paper focuses on the design of the PID temperature controller for an alkaline electrolysis system to achieve fast and stable temperature control.  A thermal dynamic model of an electrolysis system is established in the frequency-domain for controller designs. Based on this model, the temperature stability is analysed by the root distribution, and the PID parameters are optimized considering both the temperature overshoot and the settling time. The performance of the optimal PID controllers is verified through experiments. Furthermore, the simulation results show that the before-stack temperature should be used as the feedback variable for small lab-scale systems to suppress stack temperature fluctuations, and the after-stack temperature should be used for larger systems to improve the economy. This study is helpful in ensuring the temperature stability and control of electrolysis systems.
\end{abstract}

\begin{keyword}
Electrolysis system \sep temperature control \sep PID controller.
%% keywords here, in the form: keyword \sep keyword
%% MSC codes here, in the form: \MSC code \sep code
%% or \MSC[2008] code \sep code (2000 is the default)
\end{keyword}

\end{frontmatter}

\nomenclature[P]{$P$}{Electricity power}%
\nomenclature[P]{$Q$}{Thermal power}%
\nomenclature[P]{$U$}{Voltage}%
\nomenclature[P]{$I$}{Current}%
\nomenclature[P]{$T$}{Temperature}%
\nomenclature[P]{$\bar{T}$}{Average temperature}%
\nomenclature[P]{$C$}{Thermal capacity}%
\nomenclature[P]{$t$}{Time}%
\nomenclature[P]{$\tau$}{Time delay}%
\nomenclature[P]{$c$}{Specific heat capacity}%
\nomenclature[P]{$v$}{Volume flow rate}%
\nomenclature[P]{$\rho$}{Density}%
\nomenclature[P]{$k$}{Heat transfer coefficient}%
\nomenclature[P]{$A$}{Area}%
\nomenclature[P]{$R$}{Thermal resistance}%
\nomenclature[P]{$\hat{y}$}{Control signal for valve opening}%
\nomenclature[P]{$y_\mathrm{valve}$}{Valve opening}%
\nomenclature[P]{$u$}{Control variable, $u=y_\mathrm{valve}$}%
\nomenclature[P]{$T_\mathrm{f}$}{Temperature feedback}%

\nomenclature[S]{*}{Steady-state}%
\nomenclature[S]{ele}{Electrolysis}%
\nomenclature[S]{dis}{Heat dissipation}%
\nomenclature[S]{th}{Thermal neutral}%
\nomenclature[S]{sep}{Separator}%
\nomenclature[S]{c}{Cooling water}%
\nomenclature[S]{amb}{Ambient}%

\printnomenclature

%%
%% Start line numbering here if you want
%%
%%\linenumbers

%% main text
\section{Introduction}
\label{S:Introduction}
Green hydrogen, produced by renewable energy, will play a critical role in the decarbonization of the steel, chemical and transport sectors \cite{IRENA report}. As the core element of hydrogen production, it is important to ensure the safe and efficient operation of water electrolysis systems to achieve a steady hydrogen supply. However, the temperature of the electrolysis system is often disturbed by load and ambient temperature fluctuations, which affects both the system efficiency and security. Temperatures lower than the rated temperature will hinder the electrolysis reaction and lead to low efficiencies \cite{Temperature-efficiency}; on the other hand, high temperatures beyond the upper limit can harm the stack by decreasing the corrosion resistance \cite{Temperature-security}.

In existing commercial electrolysis systems, cooling devices are equipped to maintain the temperature at a set point, and PID temperature controllers are used to suppress the disturbances by regulating the cooling water flow rate \cite{Temperature control device,Second-order model}. However, heat transfer delays in electrolysis systems make PID tuning to be time-consuming, and the selected PID parameters often fail to achieve satisfactory performance. For example, stack temperature oscillations occur in \cite{Temperature oscillation} for both constant and intermittent power inputs caused by the improper PID parameter setting. In \cite{Temperature variation}, the stack temperature does not remain stable under wind power inputs, and the temperature variation is approximately 8 $^{\circ}$C with current fluctuations between 40\% to 100\% rated. Restricted by temperature variations, the electrolyte temperature is controlled at 65 $^{\circ}$C in \cite{Temperature variation}, which is far from the allowable limit of 90 $^{\circ}$C; thus, the system efficiency is sacrificed.

For the temperature control of electrolysis systems, systematic modelling and controller design methods are needed. 
Ulleberg \cite{Ulleberg} proposed a lumped model to predict the operating temperature of an advanced alkaline electrolyser. This model considers the thermal balance among the heat generation, heat loss and auxiliary cooling, which is widely used in thermal-related studies \cite{Thermal related study 1,Thermal related study 2,Thermal related study 3,Thermal related study 4,Thermal related study 5}. Y. Qiu presented an optimal production scheduling approach for utility-scale P2H plants considering the dynamic thermal process through a first-order temperature model \cite{Yiwei1}, whose parameters were estimated in \cite{Yiwei2}. The lumped model does not consider the temperature difference between the stack and the auxiliary devices. Sakas et al. \cite{Second-order model} and \cite{Third-order model} used second-order and third-order thermal models, respectively, taking into account the thermal inertia of the gas-liquid separators. 
Time delays exist in heat transfer processes and will affect the accuracy of the model when analysing the temperature of a specific component, e.g., the stack. Qi et al. \cite{Third-order model with delay} emphasized the effects of heat transfer delays on the thermal dynamic performance and added two time delay terms for the stack and the cooling coil in the third-order model.

There are few existing studies focusing on temperature control. Sakas et al. \cite{Second-order model} used a PID temperature controller in the simulation; however, the dynamic performance of the temperature controller was not discussed. Qi et al. \cite{Third-order model with delay} proposed two novel temperature controllers to reduce the temperature overshoot: a current feed-forward PID controller and a model predictive controller. As the most widely used temperature controller in commercial electrolysis systems, the tuning process of the PID temperature controller is a time-consuming task due to the multiple thermal inertia and time delay terms, which have not been discussed yet.

The focus of this paper is on the thermal dynamic analysis and  PID controller design of an alkaline electrolysis system. The main contributions are as follows.
\begin{enumerate}
	\item A frequency-domain thermal model that considers the time delays in the heat transfer process is first proposed for the controller design of electrolysis systems.

	\item The temperature stability is analysed by the root distribution, and an optimization model is proposed for parameter tuning considering both fastness and security, which is verified through experiments.
	
	\item Suggestions are given for system design to improve the thermal dynamic performance. It is suggested to use the before-stack temperature as the feedback variable for small lab-scale systems to suppress the temperature fluctuation and use the after-stack temperature for larger systems to improve the economy. In addition, time delays should be reduced to improve the thermal dynamic performance by increasing the flow rates or using shorter channels.
	
\end{enumerate}

This paper is organized as follows. In Section \uppercase\expandafter{\romannumeral2}, the complete thermal model of an alkaline electrolysis system, which is linearized and transferred to the frequency domain in Section \uppercase\expandafter{\romannumeral3} is introduced. In Section \uppercase\expandafter{\romannumeral4}, a method for PID tuning that considers the overshoot and setting time is provided.
The proposed PID tuning method is verified through experiments in Section \uppercase\expandafter{\romannumeral5}. In Section \uppercase\expandafter{\romannumeral6}, the PID temperature controllers are compared with before-stack and after-stack temperature feedbacks.
In Section \uppercase\expandafter{\romannumeral7}, the influence of time delays on the thermal dynamic performance is analysed.

\section{Temperature control of an alkaline electrolysis system by the PID temperature controller}
\subsection{System process description}
\begin{figure}[htbp]  
	\makebox[\textwidth][c]{\includegraphics[width=1\textwidth]{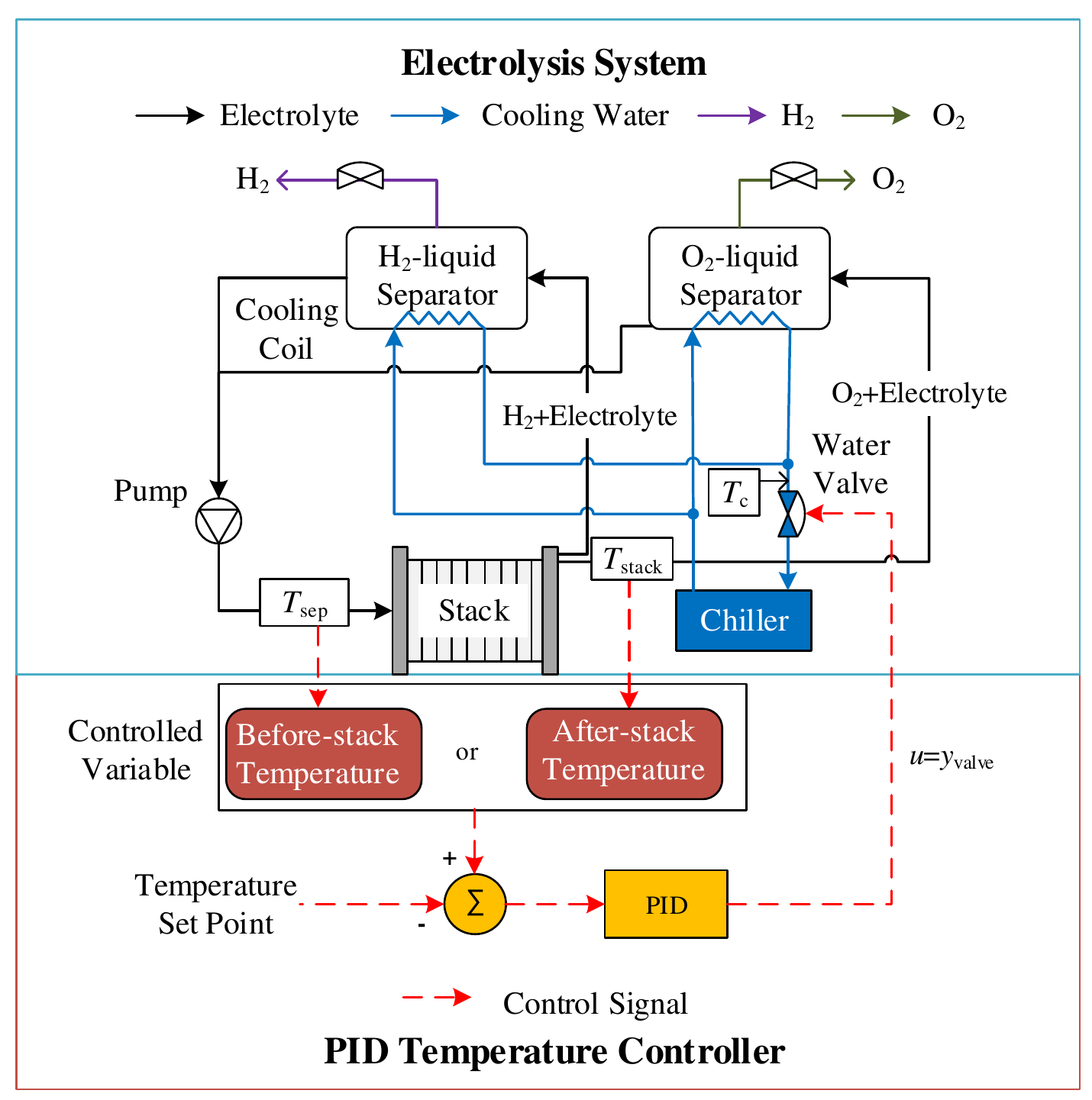}}  
	\caption{Thermal management process of an alkaline electrolysis system.}
	\label{Fig1System}
\end{figure}

The process of the analysed alkaline electrolysis system is shown in \reffig{Fig1System}. The stack is the core element of the system, in which water is electrolyzed to produce hydrogen and oxygen. The gas products mixed with electrolytes enter the gas-liquid separators, in which the gas product is separated for subsequent processing, and the remaining electrolyte from two sides are mixed and circulated into the stack.

The electrolysis reaction in the stack is exothermic at room temperature \cite{Ulleberg}. To maintain the stack temperature at the rated value, a cooling coil is placed in the gas-liquid separator to cool down the electrolytes and indirectly cool the stack. The cooling water flow rate is controlled by the water valve according to the command from the temperature controller. However, the stack temperature tends to fluctuate considerably in industrial practice due to the inappropriate parameter setting of the temperature controller as well as external disturbances, e.g., current and ambient temperature fluctuations. Thus, it becomes an important issue to obtain stable and fast temperature control.

\label{S:complete model}

\subsection{Thermal dynamic characteristics of the alkaline electrolysis system}
The thermal dynamics of the alkaline electrolysis system was modelled as a third-order with time-delays process in our previous article \cite{Third-order model with delay}. The state equations are as follows:
\begin{subequations} \label{eq:model}
	\begin{equation}\label{eq:model_stack}
	C_{\text {stack }} \frac{\mathrm{d} T_{\mathrm{stack},t}}{\mathrm{d} t}=Q_{\text {ele },t}-Q_{\mathrm{dis,stack},t}-c_\mathrm{lye}v_\mathrm{lye}\rho_\mathrm{lye}\left(T_{\mathrm{stack},t}-T_{\mathrm{sep},t-\tau_{1}}\right) 
	\end{equation}
	\begin{equation}\label{eq:model_sep}
	C_{\mathrm{sep}} \frac{\mathrm{d} T_{\mathrm{sep},t}}{\mathrm{d} t}=\frac{1}{2}v_{\mathrm{lye}} \rho_{\mathrm{lye}} c_{ \mathrm{lye}}\left(T_{\mathrm{stack},t}-T_{\mathrm{sep},t}\right)-k A \Delta T_{t}-Q_{\mathrm{dis,sep},t}
	\end{equation}
	\begin{equation}\label{eq:model_cool}
	C_{\mathrm{c}} \frac{\mathrm{d} T_{\mathrm{c},t}}{\mathrm{d} t}=v_{\mathrm{c},t-\tau_{2}} \rho_{\mathrm{c}} c_{ \mathrm{c}}\left(T_{\mathrm{c}, \mathrm{in},t}-T_{\mathrm{c},t}\right)+k A\Delta T_{t}
	\end{equation}
\end{subequations}
where $T_\mathrm{stack}$, $T_\mathrm{sep}$ and $T_\mathrm{c}$ are the temperatures of the stack, gas-liquid separator and cooling coil, respectively, as shown in \reffig{Fig1System}. Considering that $T_\mathrm{sep}$ is also the electrolyte temperature at the inlet of the stack, it is referred to as the before-stack temperature in this paper, and $T_\mathrm{stack}$ is referred to as the after-stack temperature.

In \refequ{eq:model}, there are two time delay terms $\tau_{1}$ and $\tau_{2}$. $\tau_{1}$ is the time delay of the stack, which shows that the after-stack temperature $T_\mathrm{stack}$ changes later than the before-stack temperature $T_\mathrm{sep}$ by a time-delay $\tau_{1}$ due to electrolyte convection in the stack. $\tau_{2}$ is the time delay for the cooling process. When the valve opening command $y_\mathrm{valve}$ from the controller changes, the influence is delayed by $\tau_{2}$ to the temperature of the cooling water $T_\mathrm{c}$ caused by the slow response of the cooling valve and the electrolyte convection in the cooling coil. 

\begin{figure}[htbp]  
	\makebox[\textwidth][c]{\includegraphics[width=0.9\textwidth]{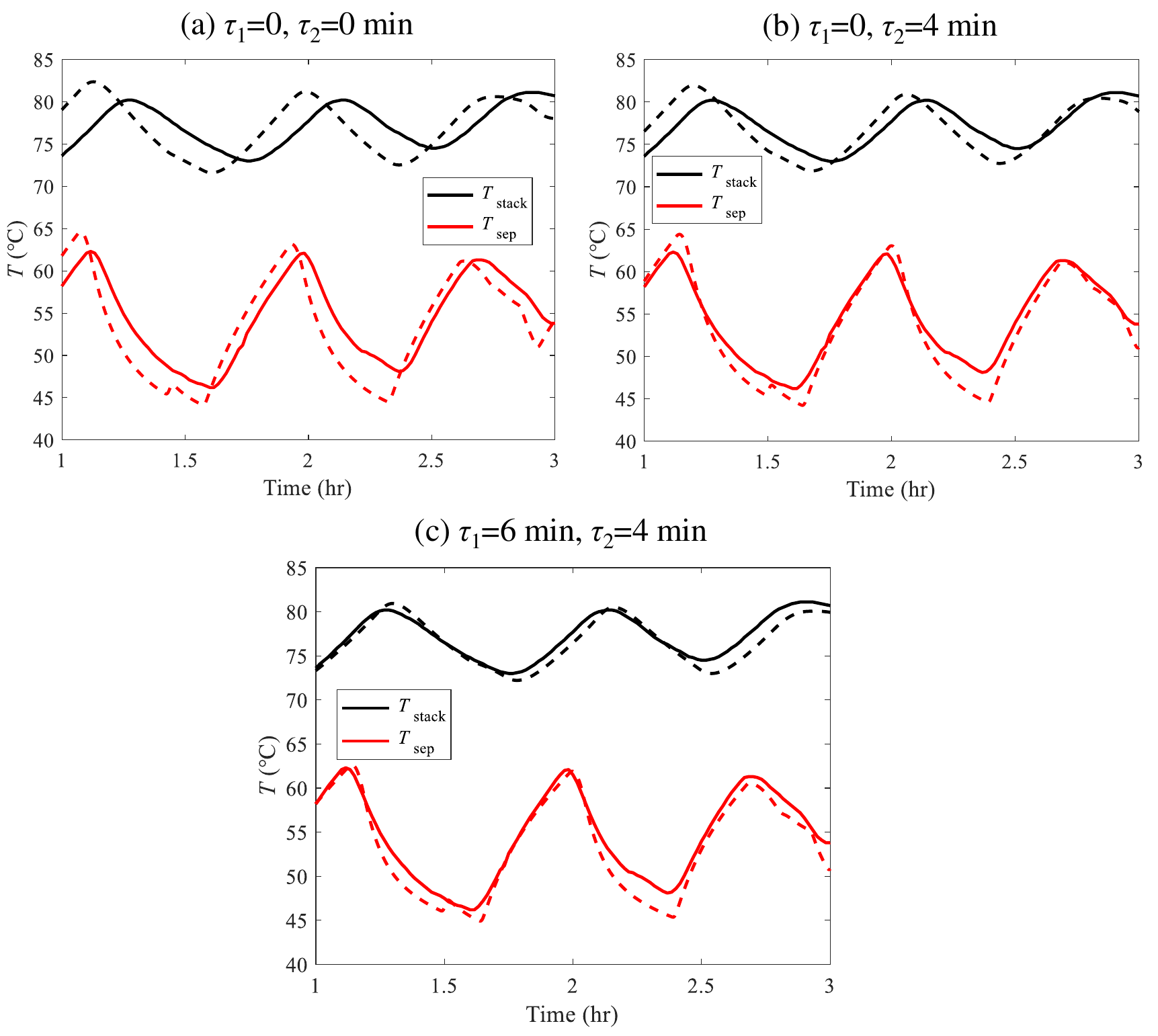}}  
	\caption{Comparison of the modelling accuracy with different time delays. The current is fixed at the rated, and the temperature oscillation is caused by improper PID parameter settings. (solid lines: experimental data; dotted lines: simulation results.)}
	\label{FigA2Delay}
\end{figure}

Our experimental results clearly show the existence of the time delays $\tau_{1}$ and $\tau_{2}$, as shown in \reffig{FigA2Delay}. The simulation results are in good agreement with the experimental data when $\tau_{1}=\SI{6}{min}$ and $\tau_{2}=\SI{4}{min}$ are adopted, which are also used in our previous paper \cite{Third-order model with delay}. These minute-long delays cause the manual tuning of the PID temperature controller to be very difficult in practice, and oscillations often occur as in \reffig{FigA2Delay}.

\subsection{PID temperature controller for the electrolysis system}
The PID temperature controller is widely used in commercial electrolysis systems \cite{Second-order model}, as follows \refequ{eq:PID}:
\begin{equation}\label{eq:PID}
y_\mathrm{valve}=k_\mathrm{p}e_{t}+k_\mathrm{i}\int e_{t}\mathrm{d} t+k_\mathrm{d}\frac{\mathrm{d} e_{t}}{\mathrm{d} t}
\end{equation}
\begin{equation}\label{eq:error}
e_{t}=T_{\mathrm{f},t}-T_{\mathrm{aim},t}
\end{equation}
where $k_\mathrm{p}$, $k_\mathrm{i}$, and $k_\mathrm{d}$ are the coefficients for the proportional, integral, and derivative terms, respectively; $e_{t}$ is the error between the temperature feedback $T_\mathrm{f}$ and the set point $T_\mathrm{aim}$ as \refequ{eq:error}, and $y_\mathrm{valve}$ is the valve opening command. 

Challenges for PID controller designs lie in both parameter tuning and selecting the feedback variable.
The PID parameters $k_\mathrm{p}$, $k_\mathrm{i}$, and $k_\mathrm{d}$ have a significant impact on the thermal dynamics of the electrolysis system. In addition, either the inlet temperature $T_\mathrm{sep}$ or the outlet temperature $T_\mathrm{stack}$ of the stack can be selected as the feedback variable for the PID temperature controller, as shown in \reffig{Fig1System}. 

This paper answers the following questions.
\begin{enumerate}
	\item How can the PID parameters be tuned to achieve stable and fast temperature control?
	\item How can the temperature feedback for the PID controller be chosen between the before-stack temperature and the after-stack temperature?
	\item Will the time delays $\tau_1$ and $\tau_2$ influence the thermal dynamic performance?
\end{enumerate}

\section{Frequency-domain model of the alkaline electrolysis system with a PID temperature controller}
In this section, the equation set \refequ{eq:model}-\refequ{eq:error} will be linearized and transformed to the frequency domain for stability analysis and controller design.
\subsection{Electrolysis system model}
The thermal dynamic model of the electrolysis system \refequ{eq:model} can be reorganized as differential-algebraic equations with delays of \refequ{eq:complete model}: 
\begin{equation}\label{eq:complete model}
\left\{\begin{array}{l}
\dot{\mathbf{x}}=\mathbf{f}\left(\mathbf{x}, \mathbf{x_{\tau}}, \mathbf{y}, \mathbf{y}_{\tau}, \mathbf{p}, u, u_{\tau_i}\right) \\
0=\mathbf{g}(\mathbf{x}, \mathbf{y}, \mathbf{p}, u) \\
0=\mathbf{g}_{i}\left(\mathbf{x}_{\tau_i}, \mathbf{y}_{\tau_i}, \mathbf{p}, u_{\tau_i}\right) \quad i=1,2.
\end{array}\right.
\end{equation}

$\mathbf{f}$ is the state equation vector consisting of the thermal dynamic model for the electrolysis system \refequ{eq:model}. 
$\mathbf{g}$ is the algebraic equation vector describing the electrochemical characteristics of the stack and the heat transfer process, as shown in \refequ{eq:Heatproduced}-\refequ{eq:Valve} in the appendix. $\mathbf{x}$ and $\mathbf{y}$ are the state variable vector and algebraic variable vector, respectively, defined as \refequ{eq:definition x} and \refequ{eq:definition y}; $u$ is the control variable as the valve opening $y_\mathrm{valve}$; $\mathbf{x_\tau}$, $\mathbf{y_\tau}$ and $u_{\tau_i}$ are the variables with time delays defined as \refequ{eq:definition xtau}-\refequ{eq:definition utau}; and $\mathbf{p}$ is the parameter vector. 
\begin{subequations}
	\begin{equation}\label{eq:definition x}
	\mathbf{x}\overset{\mathrm{def}}{=}
	\left[\begin{array}{lll}
	T_\mathrm{stack}& T_\mathrm{sep} & T_\mathrm{c}
	\end{array}\right]^\mathrm{T}
	\end{equation}
	\begin{equation}\label{eq:definition y}
	\mathbf{y}\overset{\mathrm{def}}{=}
	\left[\begin{array}{llll}
	Q_\mathrm{ele}& Q_\mathrm{dis,stack} & Q_\mathrm{dis,sep} & \Delta T_{t} 
	\end{array}\right]^\mathrm{T}
	\end{equation}
	\begin{equation}\label{eq:definition xtau}
	\mathbf{x}_{\tau_i}(t)=\mathbf{x}(t-\tau_{i}) \quad i=1,2
	\end{equation}
	\begin{equation}\label{eq:definition ytau}
	\mathbf{y}_{\tau_i}(t)=\mathbf{y}(t-\tau_{i}) \quad i=1,2
	\end{equation}
	\begin{equation}\label{eq:definition utau}
	u_{\tau_i}(t)=u(t-\tau_{i}) \quad i=1,2
	\end{equation}
\end{subequations} 

The nonlinear thermal dynamic model \refequ{eq:complete model} can be linearized and transferred to the frequency domain as \refequ{eq:linear}:
\begin{equation}\label{eq:linear}
s\mathbf{x}(s)=\mathbf{A}\mathbf{x}(s)+\mathbf{E}u(s)+\sum_{i=1}^{2}(\mathbf{A}_i \mathbf{x}_{\tau i}(s)+\mathbf{E}_{i}u_{\tau i}(s))
\end{equation}
where $\mathbf{A}$, $\mathbf{E}$, $\mathbf{A}_i$, and $\mathbf{E}_{i}$ are the Jacobian matrices, and the details can be found in \ref{S:linearization}. Then, \refequ{eq:linear} is reorganized as \refequ{eq:linear2}, and the transfer function of the electrolysis system $G_\mathrm{p}$ is derived as \refequ{eq:Gp}:
\begin{equation}\label{eq:linear2}
\mathbf{x}(s)=\frac{\mathbf{E}+\sum_{i=1}^{2}\mathbf{E}_{i}e^{-\tau_{i}s}}{s\mathbf{I}-\mathbf{A}-\sum_{i=1}^{2}\mathbf{A}_{i}e^{-\tau_{i}s}}u(s)
\end{equation}
\begin{equation}\label{eq:Gp}
G_\mathrm{p}(s)=\frac{T_\mathrm{f}(s)}{u(s)}=\frac{\mathbf{F}\mathbf{x}(s)}{u(s)}=\frac{\mathbf{F}(\mathbf{E}+\sum_{i=1}^{2}\mathbf{E}_{i}e^{-\tau_{i}s})}{s\mathbf{I}-\mathbf{A}-\sum_{i=1}^{2}\mathbf{A}_{i}e^{-\tau_{i}s}}
\end{equation}
where $\mathbf{F}$ shows the selection of the temperature feedback $T_\mathrm{f}$:
\begin{equation}\label{eq:F}
\mathbf{F}=
\begin{cases}
\begin{pmatrix}
1 & 0 & 0
\end{pmatrix},& \text{if} \quad T_\mathrm{f}=T_\mathrm{stack}, \\
\begin{pmatrix}
0 & 1 & 0
\end{pmatrix},& \text{if} \quad T_\mathrm{f}=T_\mathrm{sep}.
\end{cases}
\end{equation}

\subsection{System model with a PID temperature controller}
The PID temperature controller equations \refequ{eq:PID}-\refequ{eq:error} can also be transferred into the frequency domain:
\begin{subequations}\label{eq:PID(s)}
	\begin{equation}	u(s)=G_\mathrm{c}(s)e(s)=G_\mathrm{c}(s)(T_\mathrm{f}(s)-T_\mathrm{aim}(s))
	\end{equation}
	\begin{equation}
	G_\mathrm{c}(s)=k_\mathrm{p}+k_\mathrm{i}/s+k_\mathrm{d}s
	\end{equation}
\end{subequations}
Combining \refequ{eq:linear2} and \refequ{eq:PID(s)}, the closed-loop transfer function of the system is:
\begin{equation}\label{eq:closed-loop}
G(s)=\frac{T_\mathrm{f}(s)}{T_\mathrm{aim}(s)}=\frac{-G_\mathrm{c}(s)G_\mathrm{p}(s)}{1-G_\mathrm{c}(s)G_\mathrm{p}(s)}.
\end{equation}
The poles $\lambda$ of the closed-loop transfer function \refequ{eq:closed-loop} determine the characteristic of the thermal dynamic process, which can be calculated by the characteristic equation as \refequ{eq:pole}. The transcendental terms $e^{-\tau s}$ can be replaced by the Padé approximation, and a first-order form is \refequ{eq:Pade}.
\begin{equation}\label{eq:pole}
1-G_\mathrm{c}(\lambda)G_\mathrm{p}(\lambda)=0.
\end{equation}
\begin{equation}\label{eq:Pade}
e^{-\tau s}\approx\frac{-\tau s/2+1}{\tau s/2+1}
\end{equation}

\begin{figure}[htb]  
	\makebox[\textwidth][c]{\includegraphics[width=1.2\textwidth]{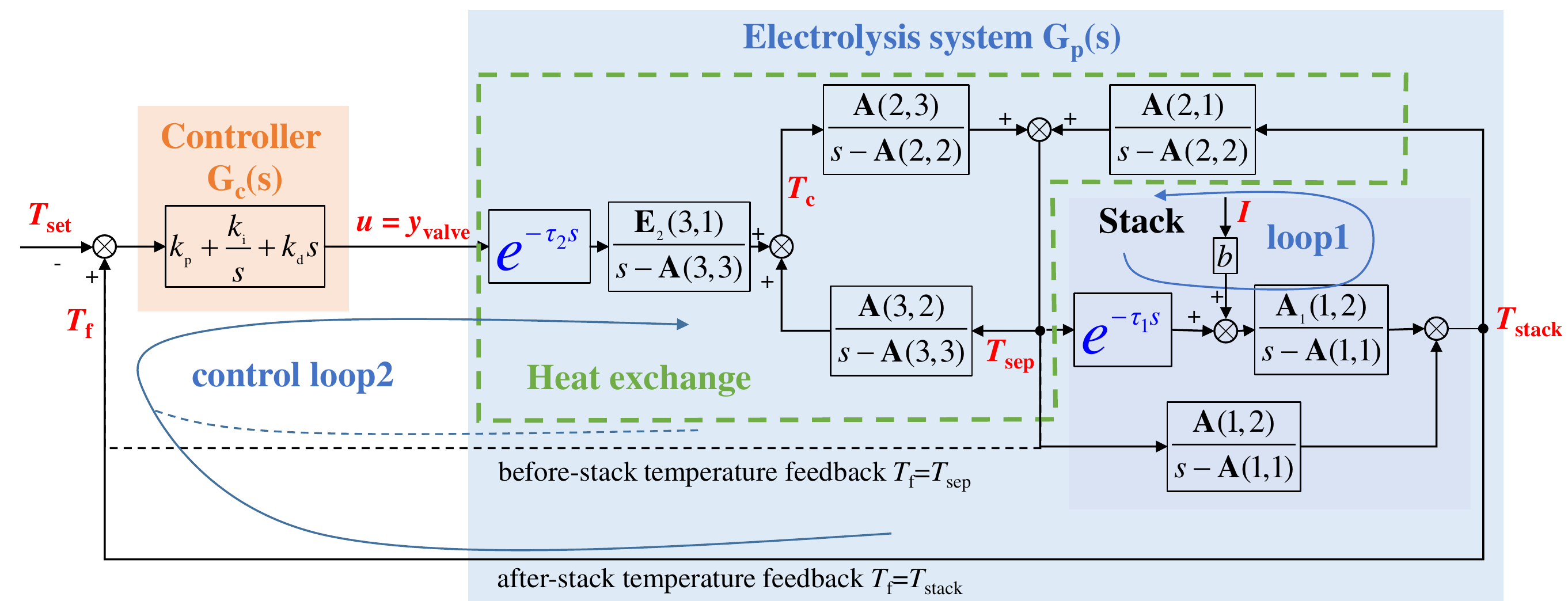}}  
	\caption{The structure block of the alkaline electrolysis system.}
	\label{FigStructureBlock}
\end{figure}

A block diagram can be derived from the linear model \refequ{eq:linear} as \reffig{FigStructureBlock} to show the thermal dynamic process more intuitively. 
\reffig{FigStructureBlock} gives a clear physical meaning to each Jacobian element in matrices $\mathbf{A}$, $\mathbf{E}$, $\mathbf{A}_i$, and $\mathbf{E}_{i}$. The diagonal elements indicate the thermal inertia of the devices. Moreover, $\mathbf{A}(1,1)$, $\mathbf{A}(2,2)$, and $\mathbf{A}(3,3)$ represent the time constant (the inverse of thermal inertia) of the stack, the gas-liquid separator and the cooling coil, respectively; while the nondiagonal elements are the gain coefficients.
Furthermore, \reffig{FigStructureBlock} compares PID temperature controllers with different temperature feedback variables $T_\mathrm{f}$, shown by dotted lines. When the after-stack temperature $T_\mathrm{stack}$ is adopted as the feedback variable $T_\mathrm{f}$, the thermal inertia of the stack, controller and separator are connected in series so that the total thermal inertia of the system increases. On the other hand, when the before-stack temperature $T_\mathrm{sep}$ is used as the feedback, the thermal inertia of the stack is in parallel with the other auxiliaries, and the system inertia is reduced, leading to better dynamic performance. The simulation results are shown in Section \ref{Discussion}.
 
\subsection{The thermal dynamic characteristics}
\begin{figure}[htbp]  
	\makebox[\textwidth][c]{\includegraphics[width=1.5\textwidth]{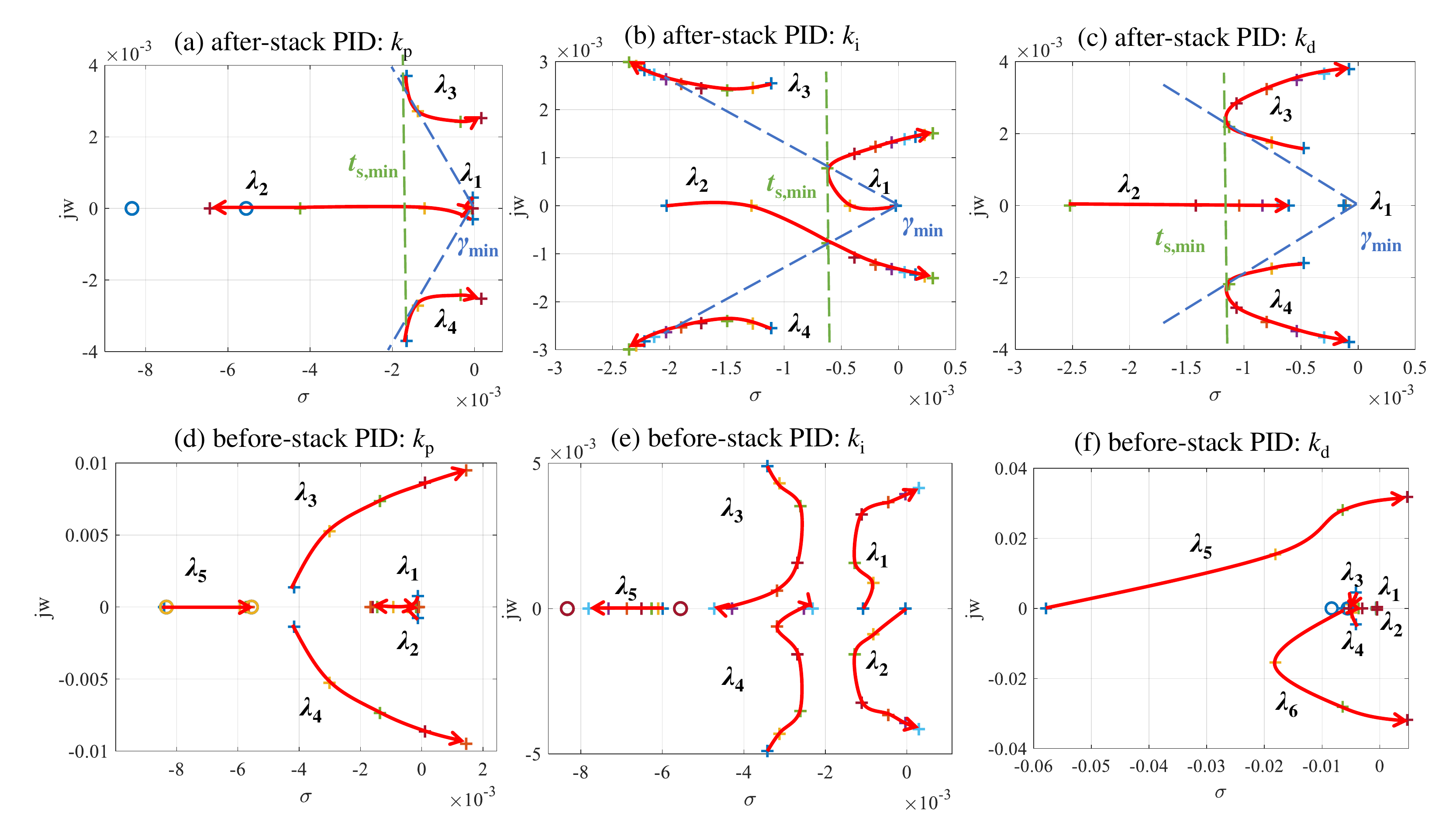}}  
	\caption{The root-locus plots with varying PID parameters (+: poles, o: zeros).}
	\label{Fig5RootLocusPlot}
\end{figure}

The closed-loop transfer function $G(s)$ \refequ{eq:closed-loop} is of high-order with $n_\lambda=6$ poles and $n_\nu=4$ zeros, in which two pole-zero pairs are introduced by the time delays $\tau$, three poles by the thermal inertia $C$ and others by the PID controller. \reffig{Fig5RootLocusPlot} shows a root-locus plot, which draws the zero and pole distributions with changing PID parameters. The parameters used for the electrolysis system model \refequ{eq:model} are from a \SI{5}{Nm^3/hr} alkaline electrolysis platform CNDQ5 \cite{Third-order model with delay} and summarized in \ref{S:Parameters}.
In \reffig{Fig5RootLocusPlot}, only the zeros and poles within 10 times the real part of the dominant pole are shown, and those far from the imaginary axis are ignored because their influences on the system dynamics are weak. The real parts of the roots are close to each other, and different poles may become the dominant pole determining the system dynamics when the PID parameters change. Therefore, there is no fixed dominant pole for the thermal dynamic process, and it is difficult to approximate it as a second-order system.

This high-order characteristic makes it difficult to use the time response-based PID tuning method, which selects the suitable pole and zero positions of a controller. Instead, we simply use an optimization model to select the optimal PID parameters considering both the temperature overshoot $\gamma$ and setting time $t_\mathrm{s}$ in the next section.

\section{Parameter tuning of the PID temperature controller}
\label{S:PID design method}
In this section, a tuning method for the PID temperature controller is provided. The stability region is derived based on the linearized frequency-domain model \refequ{eq:closed-loop}. The optimal PID parameters are selected by maximizing the temperature overshoot and the setting time in the stability region.

\subsection{Stability analysis}
The stack temperature should be asymptotically stable, which means that an equilibrium temperature is eventually reached after the disturbance. 
The stability can be judged by the stability criteria: a transfer function is stable if all its poles have a negative real part \cite{Book of control}. Therefore, the system temperature is stable if:
\begin{equation}
\lambda_i=\sigma_i+j w_i, i=1,2,\cdots, n_\lambda
\end{equation}
\begin{equation}\label{eq:stability}
\sigma_i<0, i=1,\cdots,n_\lambda.
\end{equation}
where $\sigma_i$ and $j w_i$ are the real and imaginary parts of the $i$th pole $\lambda_i$, respectively, derived from \refequ{eq:pole}. The PID parameters ($k_\mathrm{p}$, $k_\mathrm{i}$, $k_\mathrm{d}$) satisfying \refequ{eq:stability} form the stability region.

\subsection{PID tuning by optimization}
\label{section:dynamic performance}
On the premise of temperature control stability, the dynamic performance becomes important, and we pay special attention to the security and fastness of the temperature controller. The temperature overshoot $\gamma$ and the settling time $t_\mathrm{s}$ are selected as the performance indicators defined as follows:
\begin{enumerate}
	\item Temperature overshoot $\gamma$ evaluates the maximum variation of the controlled temperature from the set point $T_\mathrm{set}$ after a disturbance: 
	\begin{equation}\label{eq:overshoot}
	\gamma=(T_\mathrm{peak}-T_\mathrm{\infty})/T_\mathrm{\infty}
	\end{equation}
	where $T_\mathrm{peak}$ is the peak temperature and $T_\mathrm{\infty}$ is the steady-state value. $\gamma$ is usually defined as \refequ{eq:overshoot} under the step response process and zero initial condition in control theory \cite{Book of control}.
	A large temperature overshoot $\gamma$ will cause the stack temperature to exceed the permissible limit and cause safety problems. 
	\item The setting time $t_\mathrm{s}$ is the time required to reach and remain within $\pm$2\% of the steady state value $T_\mathrm{\infty}$ for a step response process. A long setting time $t_\mathrm{s}$ will cause temperature error accumulation since the stack temperature has not returned to the set point $T_\mathrm{set}$ when the load changes again.
\end{enumerate}

The optimal PID parameters can be derived by \refequ{eq:optimization}:
\begin{equation}\label{eq:optimization}
\underset{k_\mathrm{p},k_\mathrm{i},k_\mathrm{d}}{\min} (\gamma/\gamma_0)^2+(t_\mathrm{s}/t_\mathrm{s,0})^2
\end{equation}
where $\gamma_0$ and $t_\mathrm{s,0}$ are the reference overshoot and setting time, respectively, used to unify the order of magnitude of the two parts. The optimization problem \refequ{eq:optimization} can be solved by traversing the PID parameters in the stability region, and the overshoot $\gamma$ and setting time $t_\mathrm{s}$ are calculated by applying a unit step change on the temperature set point $T_\mathrm{aim}$ based on the closed-loop transfer function $G(s)$ as \refequ{eq:closed-loop}. 

In addition to the optimization method above, the PID parameters can also be obtained by using commercial PID parameter tuning software (e.g., MATLAB PID tuner) with the closed-loop transfer function $G(s)$ as \refequ{eq:closed-loop}. The advantage of the optimization method \refequ{eq:optimization} is that the reference values $\gamma_0$ and $t_\mathrm{s,0}$ can be easily adjusted based on the permitted temperature overshoot and setting time. In addition, an optimal PID parameter exists for the after-stack PID temperature controller at the intersection of the minimum overshoot line $\gamma_\mathrm{min}$ and the minimum setting time line $t_\mathrm{s,min}$ shown in the root-locus plot \reffig{Fig5RootLocusPlot}, which can be found by the optimization problem \refequ{eq:optimization}.

\section{PID temperature controller design and verification for a \SI{5}{Nm^3/hr} alkaline electrolysis platform}
The proposed PID tuning method is verified on a commercial \SI{5}{Nm^3/hr} alkaline electrolysis platform CNDQ5 from the Purification Equipment Research Institute of CSIC. The details of this system can be found in our previous article \cite{Third-order model with delay}. The values of the parameters in the thermal dynamic model \refequ{eq:model} are summarized in \ref{S:Parameters}.

\subsection{Optimal PID parameters}
\begin{figure}[htbp]  
	\makebox[\textwidth][c]{\includegraphics[width=1.2\textwidth]{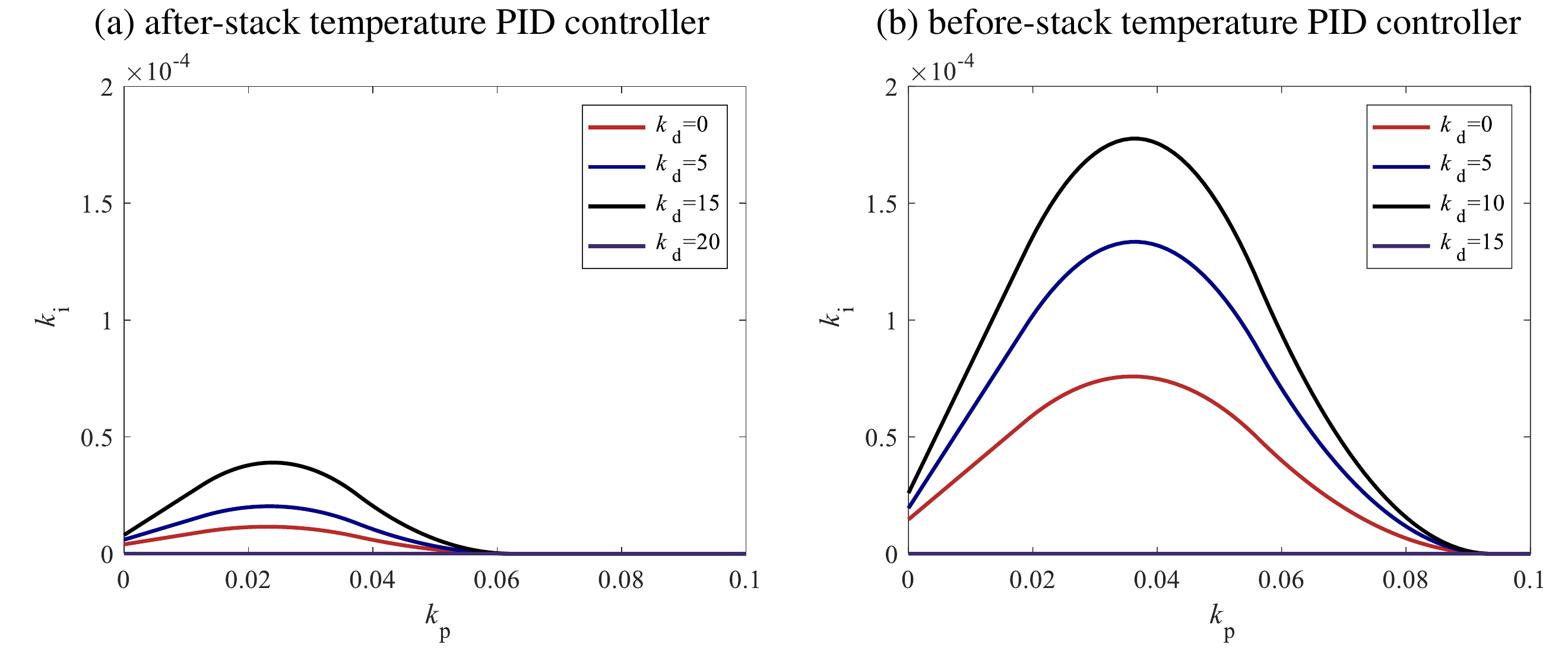}}  
	\caption{The stability regions of the after-stack and before-stack PID temperature controllers.}
	\label{FigStabilityDomain}
\end{figure}

The stability regions of the after-stack and before-stack temperature controllers are shown in \reffig{FigStabilityDomain}, which are derived by linearizing the thermal dynamic model \refequ{eq:complete model} at the equilibrium point with a rated current $I=$\SI{820}{A} and after-stack temperature $T_\mathrm{stack}=$80$^{\circ}$C. The allowed PID parameters are within the region formed by the curves and the coordinate axes. If any PID parameter is on or outside the boundary, the stack temperature $T_\mathrm{stack}$ will oscillate or overheat. Comparing the two kinds of controllers in \reffig{FigStabilityDomain}, it can be seen that the stability region of the after-stack temperature controller is much smaller than that of the before-stack temperature controller, which is why manual tuning of the after-stack temperature controller is more difficult. In addition, the after-stack temperature controller allows a larger derivative term $k_\mathrm{d}$ to predict the temperature change and offset the influence of thermal inertia and time delays. The before-stack temperature controller has a smaller upper limit for the differential term $k_\mathrm{d}$, which indicates that a PI controller may be able to reduce the parameter tuning difficulty.

The PID parameters are derived by solving the optimization problem \refequ{eq:optimization}. The reference overshoot $\gamma_0$ and setting time $t_\mathrm{s}$ are selected as 50\% and \SI{2}{hr}, respectively. A regular grid of $10\times10\times10$ sampling points of control parameters $(k_\mathrm{p},k_\mathrm{i},k_\mathrm{d})$ is employed in the traversing process.
The optimal PID parameters are shown in \reftab{tab:PIDparameters_optimal}. 

\begin{table}[htbp]
	\caption{Optimal parameters derived through optimization \refequ{eq:optimization}}
	\centering
	\begin{tabular}{cccc}
		\hline
		Type & $k_\mathrm{p}$& $k_\mathrm{i}$& $k_\mathrm{d}$\\
		\hline
		After-stack feedback& 0.02&$1.1\times10^{-5}$&6\\
		Before-stack feedback & 0.031&$3.1\times10^{-5}$&0 \\
		\hline
	\end{tabular}
	\label{tab:PIDparameters_optimal}
\end{table}

\subsection{Experimental results}
\begin{figure}[htbp]  
	\makebox[\textwidth][c]{\includegraphics[width=1.5\textwidth]{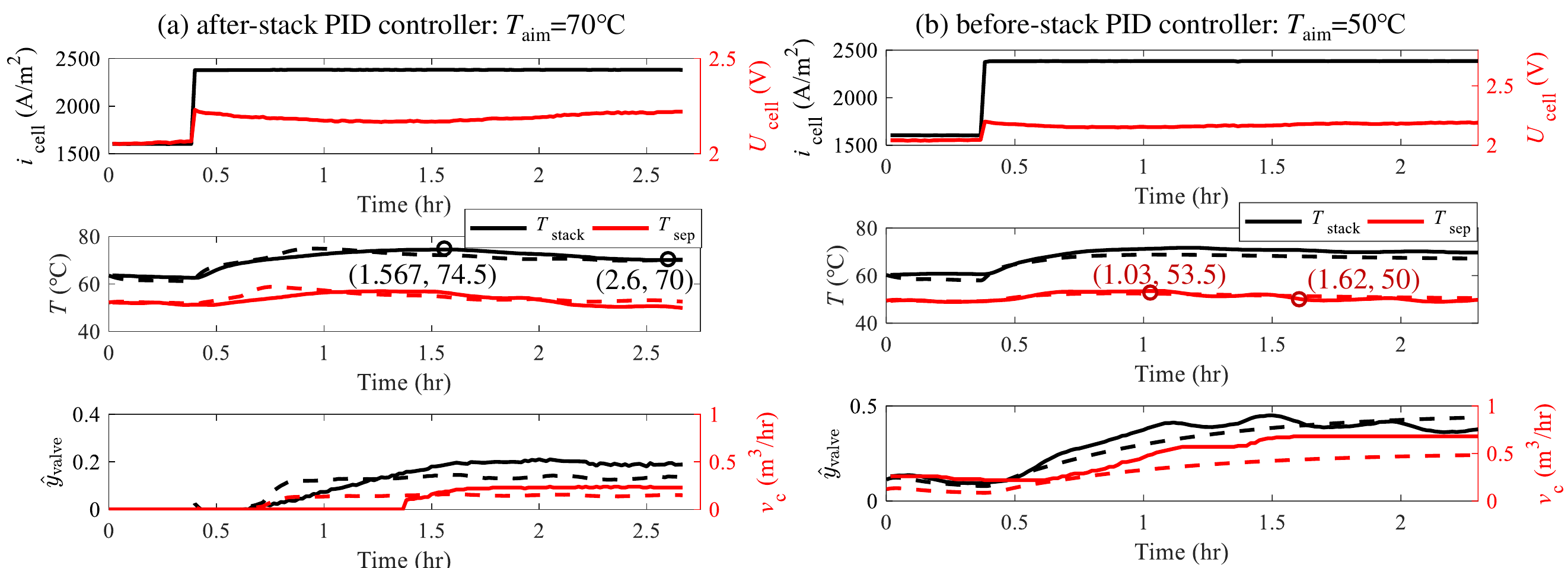}}  
	\caption{Experimental results for the optimal PID temperature controllers under a step current. (solid lines: experimental data; dotted lines: simulation results.)}
	\label{FigExperiment}
\end{figure}

The optimal PID temperature controllers in \reftab{tab:PIDparameters_optimal} are applied to the CNDQ5 platform to maintain the controlled temperature ($T_\mathrm{stack}$ or $T_\mathrm{sep}$) around the set point $T_\mathrm{aim}$ under current disturbance. The temperature set point $T_\mathrm{aim}$ of the after-stack PID controller is set to $70^{\circ}$C and the before-stack PID controller to $50^{\circ}$C.
The sampling time of the PID controller is \SI{1}{s}. The experimental results are shown in \reffig{FigExperiment}. 

It is clear that both PID controllers can achieve stable temperature control and that the controlled temperature ($T_\mathrm{stack}$ or $T_\mathrm{sep}$) gradually approaches the set point $T_\mathrm{aim}$. The oscillation that occurred in \reffig{FigExperiment}(b) is caused by the fluctuation of the cooling water inlet temperature $T_{\mathrm{c,in}}$ as a result of the improper PID parameter setting of the chiller. By using a fixed cooling water temperature $T_{\mathrm{c,in}}=30^{\circ}$C in the simulation, the oscillation is eliminated, as shown by the dotted lines in \reffig{FigExperiment}(b). In the experiment, the temperature set point $T_\mathrm{aim}$ is fixed at 70$^{\circ}$C and 50$^{\circ}$C for the after-stack PID controller and the before-stack PID controller, respectively. However, in \reffig{FigExperiment}(a), the after-stack temperature $T_\mathrm{stack}$ is lower than the set point $T_\mathrm{aim}=70^{\circ}$C during 0-\SI{0.5}{hr}, which is caused by the large heat dissipation to the ambient $Q_\mathrm{dis}$. The experiments are carried out in winter when the ambient temperature $T_\mathrm{amb}$ is approximately 0 to $10^{\circ}$C. This low temperature makes it difficult to carry out experiments. The start-up process takes approximately 4 to 5 hours. In addition, the load interval that requires cooling is very narrow, which is not conducive to testing the temperature controller. 

The simulation results of the time-domain model \refequ{eq:complete model} are shown in \reffig{FigExperiment} by the dotted lines.
The thermal dynamic model shows very high accuracy compared to the experimental data. The difference is mainly caused by the hysteretic characteristic of the cooling water valve, which is ignored in the modelling \cite{Third-order model with delay}. The high accuracy of the thermal dynamic model allows for the comparison of different PID designs and analyse the effect of time delays by simulation in the following sections.

\section{Comparison between the after-stack and before-stack PID temperature controllers}\label{Discussion}
In this section, the PID temperature controllers are compared with different feedbacks: the after-stack PID temperature controller and the before-stack PID temperature controller, shown in \reffig{Fig1System}. To obtain results with better flexibility, we use the nonlinear model \refequ{eq:complete model} in the simulation, and the ambient temperature $T_\mathrm{amb}$ is set to 25$^{\circ}$C. In addition, the optimal PID parameters in \reftab{tab:PIDparameters_optimal} are adopted.

\begin{figure}[htbp]  
	\makebox[\textwidth][c]{\includegraphics[width=1.2\textwidth]{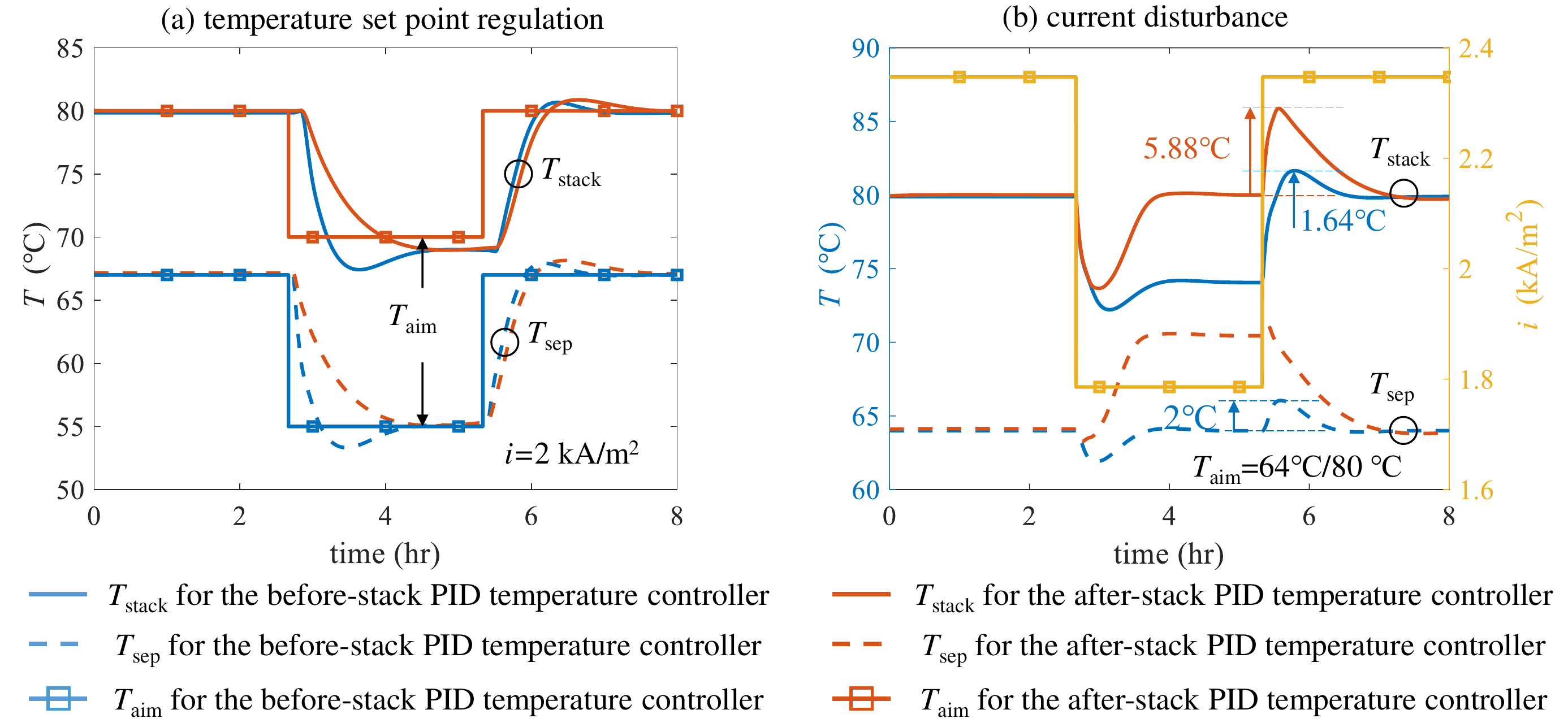}}  
	\caption{Comparison of the after-stack and before-stack PID temperature controllers under (a) temperature set point regulation and (b) current disturbance scenarios. }
	\label{Fig6Stepresponse.pdf}
\end{figure}

\reffig{Fig6Stepresponse.pdf} shows the thermal dynamic performance of the electrolysis system under the temperature set point regulation and current disturbance scenarios. In \reffig{Fig6Stepresponse.pdf}(a), it is clear that the before-stack PID temperature controller can approach the set point $T_\mathrm{aim}$ faster than the after-stack PID temperature controller. This is because the thermal inertia is reduced with a before-stack temperature feedback, as illustrated in the structure block \reffig{FigStructureBlock}. During 4.5-\SI{5.5}{hr}, the after-stack temperature $T_\mathrm{stack}$ of the system with an after-stack PID temperature controller has not reached the temperature set point $T_\mathrm{aim}=70^{\circ}$C and is still in a slow heat-up process, since the integration term $k_\mathrm{i}$ is very small at approximately $1.1\times10^{-5}$ in \reftab{tab:PIDparameters_optimal}.

In \reffig{Fig6Stepresponse.pdf}(b), the before-stack PID temperature controller also shows better resistance to current disturbances compared to the after-stack PID temperature controller. It requires a shorter time to return to the temperature set point $T_\mathrm{aim}$, and the maximum temperature deviation $\Delta T_\mathrm{max}$ is smaller: the before-stack PID temperature controller has $\Delta T_\mathrm{max}=2^{\circ}$C, and the after-stack PID temperature controller has $\Delta T_\mathrm{max}=5.88^{\circ}$C. This advantage can be explained by the structure block as \reffig{FigStructureBlock}: for the before-stack PID controller, the thermal inertia $\mathbf{A}(1,2)/(s-\mathbf{A}(1,1))$ of the stack and the heat transfer delay $\tau_{1}$ are not in the control loop 2; thus, the valve opening $y_\mathrm{valve}$ is closer to the temperature feedback $T_\mathrm{sep}$, resulting in a more timely adjustment of the cooling valve. 

\begin{figure}[htbp]  
	\makebox[\textwidth][c]{\includegraphics[width=0.7\textwidth]{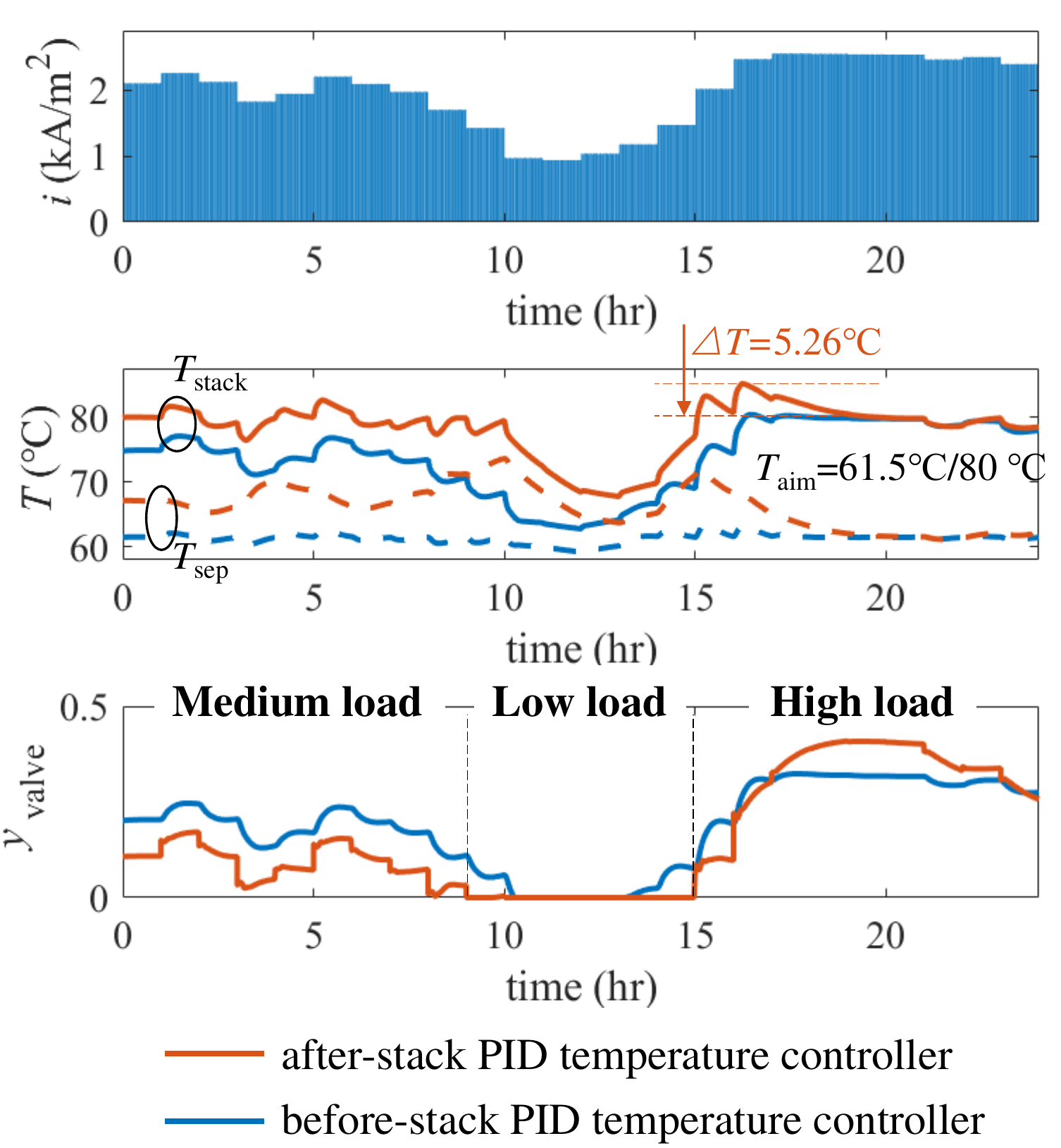}}  
	\caption{Comparison of the after-stack and before-stack PID temperature controllers under the peak shaving scenario.}
	\label{Fig7PeakRegulation}
\end{figure}

The resistance to current disturbances is significant for electrolysis systems under dynamic operation, e.g., in renewable-to-hydrogen scenarios. The thermal dynamic performance of the two PID temperature controllers are compared in \reffig{Fig7PeakRegulation} with the wind power input from a wind turbine generator that is averaged and received every \SI{1}{hr} \cite{Wind data}. Statistics indicating the temperature control performance are shown in \reftab{tab:peakshaving}. 

From \reftab{tab:peakshaving}, the before-stack PID temperature controller has better dynamic performance considering the maximum temperature deviation $\Delta T_\mathrm{max}$ and the fluctuation of the controlled temperature $\delta_T$. For the after-stack PID temperature controller, the stack temperature $T_\mathrm{stack}$ exceeds the set point $T_\mathrm{aim}$ with a deviation of $\Delta T_\mathrm{max}=5.26^{\circ}$C when the current suddenly increases at $t=$\SI{16}{hr}. This temperature deviation $\Delta T_\mathrm{max}$ makes it necessary for the after-stack PID temperature controller to leave a certain margin between the temperature set point $T_\mathrm{aim}$ and the upper limit; otherwise, the stack will be overheated and cause safety problems. Furthermore, the controlled temperature fluctuation of the after-stack PID temperature controller in the medium load region is larger than that of the before-stack PID temperature controller, as shown in $\delta_T$ in \reftab{tab:peakshaving}. This can also be observed in \reffig{Fig7PeakRegulation} since the before-stack temperature $T_\mathrm{sep}$ of the before-stack PID temperature controller is stable in the full operating intervals.

\begin{table}[htbp]
	\caption{Comparison of the thermal dynamic performance in peak shaving scenario}
	\label{tab:peakshaving}
	\centering
	\begin{threeparttable}
	\begin{adjustbox}{center}
		\begin{tabular}{ccc}
			\hline
			& Before-stack PID & After-stack PID\\
			\hline
			\tabincell{c}{Maximum temperature deviation\\ from the set point $\Delta T_\mathrm{max}$}& 1.82$^{\circ}$C &5.26$^{\circ}$C \\
			\tabincell{c}{Controlled temperature variance\\ during medium loading $\delta_T$*}& 0.46$^{\circ}$C & 1.31$^{\circ}$C\\ 
			Average stack temperature $\bar{T}$**& 67.58$^{\circ}$C & 72.27$^{\circ}$C \\
			\hline
		\end{tabular}
	\end{adjustbox}
\begin{tablenotes}
	\item[*] $\delta_T=\sqrt{\sum_{i=1}^{N}(T_i-T_{\mathrm{aim},i})^2/N}$, $T=T_\mathrm{sep}$ or $T_\mathrm{stack}$
	\item[**] 
	$\bar{T}=\sum_{i=1}^{N}(T_{\mathrm{sep},i}+T_{\mathrm{stack},i})/(2N)$
\end{tablenotes}

\end{threeparttable}
\end{table}

Although the before-stack PID temperature controller has the advantages described above, it is not always the best choice. The average stack temperature $\bar{T}$ is reduced during low and medium loading periods since the before-stack temperature $T_\mathrm{sep}$ is fixed at $T_\mathrm{set}$, leading to an efficiency loss. On the other hand, the after-stack PID temperature controller can achieve a high average electrolysis temperature $\bar{T}$ and ensure safe operation since the hottest spot in the stack is always at the outlet $T_\mathrm{stack}$.

\reftab{tab:PIDcontroller} gives suggestions on the temperature feedback selection. For small lab-scale systems, it is preferred to use the before-stack temperature $T_\mathrm{sep}$ as the feedback because the thermal capacity $C$ of the system is small and the stack temperature tends to fluctuate considerably under disturbances, e.g., current and ambient temperature. In addition, the after-stack PID temperature controller is more suitable for larger commercial systems to improve the economy.

\begin{table}[htbp]
	\caption{PID controller comparison}
	\label{tab:PIDcontroller}
	\centering
	\begin{adjustbox}{center}
		\begin{tabular}{ccc}
			\hline
			Controller& Advantages & Disadvantages\\
			\hline
			Before-stack PID& \tabincell{c}{Less fluctuation under \\dynamic operation} & \tabincell{c}{Low average temperature\\ during low-load periods}\\
			After-stack PID& \tabincell{c}{High efficiency} & \tabincell{c}{More fluctuation\\ under dynamic operation} \\
			\hline
		\end{tabular}
	\end{adjustbox}
\end{table}

\section{Effect of the time delays on the thermal dynamic performance}
The impact of the time delays $\tau_1$ and $\tau_2$ on the thermal dynamic performance is analysed in this section to optimize the performance in the system design process.

\begin{figure}[htbp]  
	\makebox[\textwidth][c]{\includegraphics[width=1.2\textwidth]{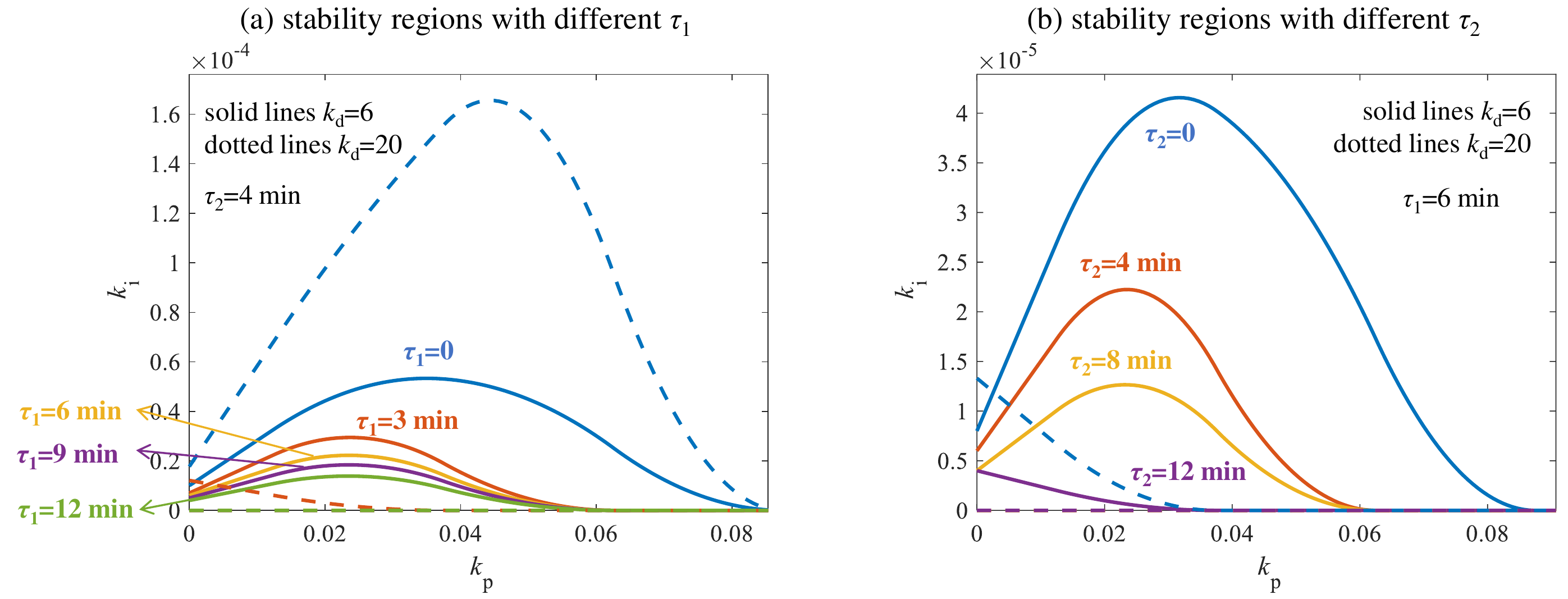}}  
	\caption{Stability region of the after-stack PID controller at different time-delays.}
	\label{StabilitywithDelay_afterstack}
\end{figure}

The stability regions with different time delays for the after-stack PID temperature controller are shown in \reffig{StabilitywithDelay_afterstack}. When the time delay $\tau_{1}$ or $\tau_{2}$ increases, the stability region decreases monotonically. This indicates that for the after-stack PID temperature controller with large time delays, the allowed PID parameter domain is extremely narrow, which causes PID tuning to be difficult.

\begin{figure}[htbp]  
	\makebox[\textwidth][c]{\includegraphics[width=1.2\textwidth]{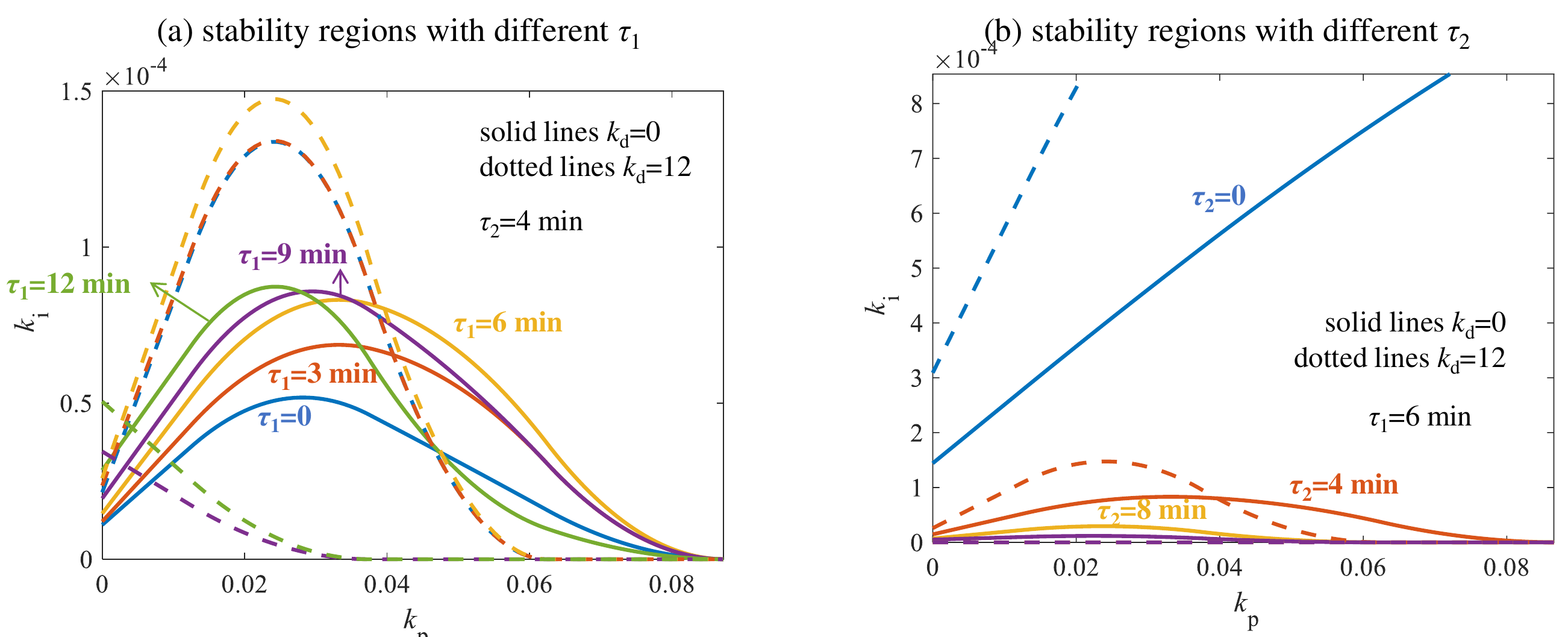}}  
	\caption{Stability region of the before-stack PID controller at different time-delays.}
	\label{StabilitywithDelay_beforestack}
\end{figure}

For the before-stack PID temperature controller, the influence of the time delay of the stack $\tau_{1}$ on the stability region is different from that of the after-stack PID controller, as shown in \reffig{StabilitywithDelay_beforestack}(a). 
When the derivative term $k_\mathrm{d}=0$, the stability region increases slightly and then remains almost the same with the increasing time delay of the stack $\tau_{1}$. This is attributed to the position of $\tau_{1}$, as shown in the structure block \reffig{FigStructureBlock}. The time delay of the stack $\tau_{1}$ is in the positive feedback loop 1, which can be beneficial to the system stability for some PID parameter settings. For example, when the before-stack temperature $T_\mathrm{sep}$ increases due to disturbances, $\tau_{1}$ delays the further increase of $T_\mathrm{sep}$ through loop 1.
On the other hand, the time delay of the cooling process $\tau_{2}$ is inside the control loop 2, which is a negative feedback loop and therefore worsens the stability, as shown in \reffig{StabilitywithDelay_beforestack}(b). 

\begin{figure}[htbp]  
	\makebox[\textwidth][c]{\includegraphics[width=1.2\textwidth]{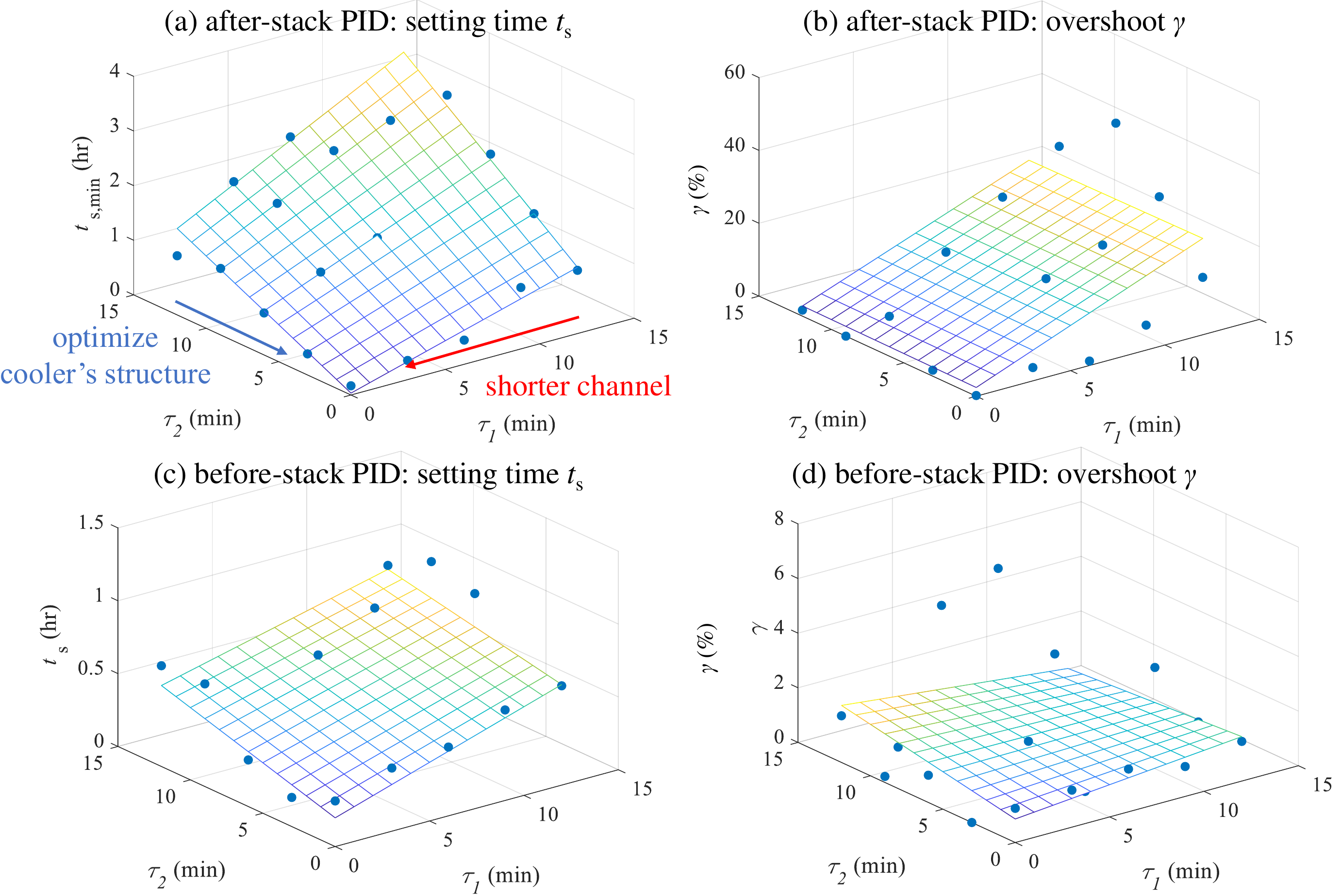}}  
	\caption{Setting time $t_\mathrm{s}$ and overshoot $\gamma$ at different time-delays. (a)(b) After-stack PID temperature controller. (c)(d) Before-stack PID temperature controller.}
	\label{Delay}
\end{figure}

The relationships between the time delays and the thermal dynamic performance of the system are analysed using performance indicators, that is, the temperature overshoot $\gamma$ and the settling time $t_\mathrm{s}$, shown as \reffig{Delay}. The blue dots show the performances of the optimal PID controllers at the different time delays $\tau_{1}$ and $\tau_{2}$. A surface as \refequ{eq:fitting} is used to show the changing trend, in which $a_1$-$a_4$ are the fitting parameters.
\begin{equation}\label{eq:fitting}
\tau/t_\mathrm{s}=a_1+a_2\tau_{1}+a_3\tau_{2}+a_4\tau_{1}\tau_{2}
\end{equation}

From \reffig{Delay}, it is clear that smaller time delays $\tau_{1}$$\tau_{2}$ lead to faster response $t_\mathrm{s}$ and reduced overshoot $\gamma$, which is beneficial to effective temperature control. Compared with the no-delay circumstance, $\tau_{1}=\tau_{2}=\SI{12}{min}$ will increase the setting time $t_\mathrm{s}$ from \SI{0.17}{hr} to \SI{3.69}{hr} and the overshoot $\gamma$ from $0\%$ to $25.94\%$ for the after-stack PID controller. A similar phenomenon is also observed for the before-stack PID controller: $\tau_{1}=\tau_{2}=\SI{12}{min}$ increases the setting time $t_\mathrm{s}$ from \SI{0.197}{hr} to \SI{0.96}{hr}, while the overshoot $\gamma$ remains 2\% below. Furthermore, from the slope of the surface, it can be seen that the before-stack PID temperature controller is less sensitive to the time delay variation and can tolerate large time delays. For the before-stack PID temperature controller, $\tau_{2}$ has a greater impact on the performance compared to $\tau_{1}$, which is consistent with the results obtained from the stability analysis in \reffig{StabilitywithDelay_beforestack}.

Time delays should be reduced during the system design to avoid slow temperature regulations and large overshoots. The time delay of the stack $\tau_1$ is caused by the electrolyte convection in the stack and can be reduced by increasing the electrolyte flow rate or using shorter channels. The time delay of the cooling process $\tau_{2}$ represents the time from the change in the valve opening command $y_\mathrm{valve}$ to the change in the outlet temperature of the cooling coil $T_\mathrm{c}$, which can be reduced by adopting a larger cooling water flow rate $v_\mathrm{c}$ or optimizing the structure of the cooler (shorter cooling pipelines). In addition, moving the cooler closer to the stack, rather than putting the cooling coil in the gas-liquid separator as \reffig{Fig1System}, is also beneficial to improve the thermal performance because the cooling effect can be applied on the stack as soon as possible. 

In recent years, large-scale alkaline electrolysis systems have become a development trend, and the long pipelines between modules make the time-delays obvious. In such scenarios, evaluating and eliminating the influence of time delays become significant to improve the dynamic performance of the system.

\section{Conclusion}
This study shows how to design a PID temperature controller in an alkaline electrolysis system with time delays. A frequency-domain model \refequ{eq:closed-loop} that can be directly used for PID tuning is derived. In addition, an optimization-based PID tuning method is proposed to minimize both the temperature setting time $t_\mathrm{s}$ and overshoot $\gamma$.

The effectiveness of the proposed PID tuning method is verified through experiments. Furthermore, a detailed comparison between the after-stack and the before-stack PID temperature controllers is carried out through simulation. The results show that the before-stack PID temperature controller has better thermal dynamic performance than the after-stack PID temperature controller, although the system efficiency is sacrificed at low-loading periods. It is suggested to use the before-stack temperature $T_\mathrm{sep}$ as the feedback variable for small lab-scale systems to suppress the stack temperature fluctuation and use the after-stack temperature $T_\mathrm{stack}$ for larger systems to improve the economy.

The influence of time delays on the thermal dynamic performance is also discussed. Larger time delays result in a longer setting time $t_\mathrm{s}$ and larger overshoot $\gamma$. Optimization methods should be considered in the design process, including increasing the electrolyte and cooling water flow rate, optimizing the structure of the cooler, and moving the cooler closer to the stack.

This study is helpful for the temperature controller design of electrolysis systems. Although the alkaline electrolysis system is the focus in this study, the method given is suitable for both alkaline and proton exchange membrane (PEM) electrolysis systems and for different system structures.

%\section*{Acknowledgement}

%% The Appendices part is started with the command \appendix;
%% appendix sections are then done as normal sections
\appendix

\section{Algebraic equations in the thermal dynamic model}
 The heat produced $Q_{\mathrm{ele},t}$ is proportional to the electrolysis current $I_\mathrm{cell}$ as \refequ{eq:Heatproduced}: 
\begin{equation}\label{eq:Heatproduced}
Q_\mathrm{ele}=(U_\mathrm{cell}-U_\mathrm{th}) I_\mathrm{cell}N_\mathrm{cell}
\end{equation}
and an empirical relationship is used for calculating the cell voltage $U_\mathrm{cell}$:
\begin{equation}\label{eq:UIcurve}
U_{\text {cell },t}=U_{\text {rev }}+(r_{1}+r_{2} \bar{T}_t)i_t+s \log \left((t_{1}+t_{2} / \bar{T}_t+t_{3} / \bar{T}_t^{2})i_t +1\right)
\end{equation}
\begin{equation}\label{eq:average temperature}
\bar{T}_{t}=(T_{\mathrm{stack},t}+T_{\mathrm{sep},t})/2.
\end{equation}
The heat loss to the ambient $Q_\mathrm{dis,stack}$ is composed of the thermal convection term $Q_\mathrm{conv}$ and the radiation term $Q_\mathrm{rad}$:
\begin{equation}
Q_\mathrm{dis,stack}=Q_\mathrm{conv}+Q_\mathrm{rad}=h A_\mathrm{stack}(T_\mathrm{stack}-T_\mathrm{amb})+\sigma A_\mathrm{stack} \varepsilon_\mathrm{stack}\left(T_\mathrm{stack}^{4}-T_\mathrm{amb}^{4}\right)
\end{equation}
\begin{equation}
h=2.51 \times 0.52\left(\frac{(T_\mathrm{stack}-T_\mathrm{amb})}{\varphi_\mathrm{stack}}\right)^{0.25}.
\end{equation}
The time-delay term $\tau_{1}$ in \refequ{eq:model_stack} represents the delay from the change in the before-stack temperature $T_\mathrm{sep}$ to the change in the after-stack temperature $T_\mathrm{stack}$.

\refequ{eq:model_sep} and \refequ{eq:model_cool} illustrate the heat exchange process from the electrolyte in the separator to the cooling water. $T_\mathrm{sep}$ is the outlet temperature of the separator, which is also the before-stack temperature, as shown in \reffig{Fig1System}, and $T_\mathrm{c}$ is the outlet temperature of the cooling coil. The mean logarithmic temperature difference $\Delta T_t$ and the heat dissipation to the ambient $Q_\mathrm{dis,sep}$ are:
\begin{equation}
\Delta T_t=\frac{\left(T_{\mathrm{stack},t}-T_{\mathrm{c},t}\right)-\left(T_{\mathrm{sep},t}-T_{\mathrm{c}, \mathrm{in},t}\right)}{\ln \left(\left(T_{\mathrm{stack},t}-T_{\mathrm{c},t}\right) /\left(T_{\mathrm{sep},t}-T_{\mathrm{c}, \mathrm{in},t}\right)\right)}
\end{equation}
\begin{equation}
Q_\mathrm{dis,sep}=\frac{\bar{T}-T_\mathrm{amb}}{R_\mathrm{sep}}.
\end{equation}
The cooling water flow rate $v_{\mathrm{c},t}$ in \refequ{eq:model_cool} is  
proportional to the valve opening $y_{\mathrm{valve},t}$ as \refequ{eq:Valve}, and a time-delay term $\tau_{2}$ is introduced to show the delay from the change of cooling water flow rate $v_\mathrm{c}$ to the cooling coil's temperature $T_\mathrm{c}$.
\begin{equation}\label{eq:Valve}
v_{\mathrm{c},t}=k_\mathrm{valve}y_{\mathrm{valve},t}.
\end{equation}

\section{Methods for model linearization and transformation to frequency domain}\label{S:linearization}
Discretizing the nonlinear model \refequ{eq:complete model} at an equilibrium point ($\mathbf{x}^*$, $\mathbf{y}^*$, $u^*$):
\begin{equation}\label{eq:discrete model}
\left\{\begin{array}{l}
\Delta \dot{\mathbf{x}}=\tilde{\mathbf{A}} \Delta \mathbf{x}+\tilde{\mathbf{B}} \Delta \mathbf{y}+
\sum_{i=1}^{2}\left(\tilde{\mathbf{A}}_{i} \Delta \mathbf{x}_{\tau_i}+\tilde{\mathbf{B}}_{i} \Delta \mathbf{y}_{\tau_ i}\right) \\
0=\tilde{\mathbf{C}} \Delta \mathbf{x}+\tilde{\mathbf{D}} \Delta \mathbf{y}+\tilde{\mathbf{E}}\Delta u \\
0=\tilde{\mathbf{C}}_{i} \Delta \mathbf{x}_{\tau_i}+\tilde{\mathbf{D}}_{i} \Delta \mathbf{y}_{\tau_i}+\tilde{\mathbf{E}}_{i} \Delta u_{\tau_{i}}, \quad i=1,2
\end{array}\right.
\end{equation}
where $\Delta \mathbf{x}=\mathbf{x}-\mathbf{x^*}$ and $\Delta \mathbf{y}=\mathbf{y}-\mathbf{y^*}$, and the Jacobian matrices are as follows:
\begin{equation}
\tilde{\mathbf{A}}=J(\mathbf{f},\mathbf{x}) \quad \tilde{\mathbf{B}}=J(\mathbf{f},\mathbf{y}) \quad \tilde{\mathbf{C}}=J(\mathbf{g},\mathbf{x}) \quad \tilde{\mathbf{D}}=J(\mathbf{g},\mathbf{y}) \quad
\tilde{\mathbf{E}}=J(\mathbf{g},\mathbf{u})
\end{equation}
\begin{equation}
\tilde{\mathbf{A}_i}=J(\mathbf{f},\mathbf{x}_{\tau_i}) \quad \tilde{\mathbf{B}_i}=J(\mathbf{f},\mathbf{y}_{\tau_i}) \quad \tilde{\mathbf{C}_i}=J(\mathbf{g}_i,\mathbf{x}_{\tau_i}) \quad \tilde{\mathbf{D}_i}=J(\mathbf{g}_i,\mathbf{y}_{\tau_i}) \quad
\tilde{\mathbf{E}_i}=J(\mathbf{g}_i,\mathbf{u}_{\tau_i})
\end{equation}
\begin{equation}
J(\mathbf{f},\mathbf{x})=\left[\begin{array}{ccc}
\frac{\partial f_{1}}{\partial x_{1}} & \cdots & \frac{\partial f_{1}}{\partial x_{n}} \\
\vdots & \ddots & \vdots \\
\frac{\partial f_{n}}{\partial f_{1}} & \cdots & \frac{\partial f_{n}}{\partial x_{n}}
\end{array}\right]_{\mathbf{x}=\mathbf{x^*}}
\end{equation}

Then, $\Delta \mathbf{y}$ and $\Delta \mathbf{y}_i$ are eliminated from \refequ{eq:discrete model} to obtain a simplified form:
\begin{equation}\label{eq:simplified model}
\begin{aligned}
\Delta \dot{\mathbf{x}} &=\tilde{\mathbf{A}} \Delta \mathbf{x}-\tilde{\mathbf{B}} \tilde{\mathbf{D}}^{-1} (\tilde{\mathbf{C}} \Delta \mathbf{x}+\tilde{\mathbf{E}}\Delta u)+\sum_{i=1}^{2}\left(\tilde{\mathbf{A}}_{i} \Delta \mathbf{x}_{\tau_i}-\tilde{\mathbf{B}}_{i} \tilde{\mathbf{D}}_{i}^{-1} (\tilde{\mathbf{C}}_{i} \Delta \mathbf{x}_{\tau_i}+\tilde{\mathbf{E}}_{i} \Delta u_{\tau_{i}})\right) \\
&=\mathbf{A} \Delta \mathbf{x}+\sum_{i=1}^{2} \mathbf{A}_{i} \Delta \mathbf{x}_{\tau_i}+\mathbf{E}\Delta u+ \sum_{i=1}^{2} \mathbf{E}_{i} \Delta u_{\tau_{i}}
\end{aligned}
\end{equation}
where $\mathbf{A}$, $\mathbf{E}$ are the reduced-order Jacobian matrices and $\mathbf{A}_i$, $\mathbf{E}_i$ are the reduced-order Jacobian matrices with delay:
\begin{equation}
\mathbf{A}=\tilde{\mathbf{A}}-\tilde{\mathbf{B}} \tilde{\mathbf{D}}^{-1} \tilde{\mathbf{C}}
\end{equation}
\begin{equation}
\mathbf{A}_{i}=\widetilde{\mathbf{A}}_{i}-\widetilde{\mathbf{B}}_{i} \tilde{\mathbf{D}}_{i}^{-1} \tilde{\mathbf{C}}_{i}
\end{equation}
\begin{equation}
\mathbf{E}=-\tilde{\mathbf{B}} \tilde{\mathbf{D}}^{-1} \tilde{\mathbf{E}}
\end{equation}
\begin{equation}
\mathbf{E}_{i}=-\widetilde{\mathbf{B}}_{i} \tilde{\mathbf{D}}_{i}^{-1} \tilde{\mathbf{E}}_{i}.
\end{equation}

The linear model \refequ{eq:simplified model} is transferred into the frequency domain by Laplace transform to get \refequ{eq:linear}.

\section{Parameters for the thermal dynamic model}
\label{S:Parameters}
Thermally related parameters for the \SI{5}{m^3/hr} alkaline electrolysis systems are shown in \reftab{tab: 50AEL}, and the parameters for the cell's U-I curve \refequ{eq:UIcurve} are shown in \reftab{tab:UIcurve}.
\begin{table}[htbp]
	\caption{Parameters for the alkaline electrolysis systems}
	\centering
	\begin{threeparttable}
		\begin{tabular}{cc}
			\hline
			Parameters & \SI{5}{Nm^3/hr}\\
			\hline
			Cell number $N_\mathrm{cell}$& 26 (13 cells in series)\\
			Cell diameter & \SI{0.5}{m}\\ 
			Cell area $A_\mathrm{cell}$ & \SI{0.196}{m^2}\\
			Stack diameter $\varphi_\mathrm{stack}$ & \SI{0.61}{m}\\
			Stack length $L_\mathrm{stack}$& \SI{0.267}{m}\\
			Stack surface area $A_\mathrm{stack}$& \SI{1.1}{m^2}\\
			Electrode volume $V_\mathrm{stack,electrode}$ & \SI{0.03}{m^3}\\
			Free stack volume $V_\mathrm{stack,free}$& \SI{0.05}{m^3}\\
			Stack void fraction at rated $f_\mathrm{v}$ &0.5\\
			Blackness of the stack surface $\varepsilon_\mathrm{stack}$ & 0.8\\
			Separator diameter $\varphi_\mathrm{sep}$ & \SI{0.219}{m}\\
			Separator length $L_\mathrm{sep}$ & \SI{2}{m}\\
			Separator volume $V_\mathrm{sep}$ & \SI{1.38}{m^3}\\
			Separator liquid level $h_\mathrm{l,sep}$& 50\% \\
			Heat transfer coefficient of the cooling coil $kA$ &\SI{140}{W/K}\\
			Lye composition & KOH\\
			Mass fraction of KOH in electrolyte $w_\mathrm{lye}$ & 31.2\%\\
			Electrolyte flow rate $v_\mathrm{lye}$&0.4-\SI{0.6}{m^3/hr}\\
			Thermal resistance $R_\mathrm{sep}$ & \SI{0.04}{K/W}\\
			Stack heat capacity $C_\mathrm{stack}$& \SI{120}{KJ/K}\\
			Separator heat capacity $C_\mathrm{sep}$ &\SI{146}{KJ/K}\\
			Cooling coil heat capacity $C_\mathrm{c}$&\SI{23}{KJ/K}\\
			Stack time-delay $\tau_{1}$ & \SI{6}{min}\\
			Cooling coil time-delay $\tau_{2}$ & \SI{4}{min}\\
			\hline
		\end{tabular}
	\end{threeparttable}
	\label{tab: 50AEL}
\end{table}

\begin{table}[htbp]
	\caption{U-I curve parameters}
	\centering
	\begin{tabular}{cc}
		\hline
		Parameters & Values \\
		\hline
		$r_1$  & \SI{1.71e-4}{\ohm m^2} \\
		$r_2$ & \SI{-1.96e-7}{\ohm m^2/K}\\ 
		$s$ & \SI{0.16}{V}\\
		$t_1$ & \SI{-0.24}{m^2/A}\\
		$t_2$ & \SI{26.23}{m^2K/A}\\
		$t_3$ & \SI{139.88}{m^2K^2/A}\\
		\hline
	\end{tabular}
	\label{tab:UIcurve}
\end{table}

%% \section{}
%% \label{}

%% References
%%
%% Following citation commands can be used in the body text:
%% Usage of \cite is as follows:
%%   \cite{key}          ==>>  [#]
%%   \cite[chap. 2]{key} ==>>  [#, chap. 2]
%%   \citet{key}         ==>>  Author [#]

%% References with bibTeX database:

% \bibliographystyle{model1-num-names}

%% New version of the num-names style
%\bibliographystyle{elsarticle-num-names}
%\bibliography{sample.bib}

%% Authors are advised to submit their bibtex database files. They are
%% requested to list a bibtex style file in the manuscript if they do
%% not want to use model1-num-names.bst.

%% References without bibTeX database:

\end{document}